\documentclass[11pt]{article}

\usepackage{amsmath,amssymb,mathtools,amsfonts,epsfig,graphicx,euscript}
\usepackage{amsmath,amssymb}
\newcommand{\bse}{\begin{subequations}}
\newcommand{\ese}{\end{subequations}}
\newcommand{\be}{\begin{equation}}
\newcommand{\ee}{\end{equation}}
\newcommand{\bea}{\begin{eqnarray}}
\newcommand{\eea}{\end{eqnarray}}
\newcommand{\ba}{\begin{array}}
\newcommand{\ea}{\end{array}}

\newcommand{\eq}[1]{(\ref{#1})}

\begin{document}
\begin{titlepage}
\thispagestyle{empty}

\vspace{2cm}
\begin{center}
\font\titlerm=cmr10 scaled\magstep4 \font\titlei=cmmi10
scaled\magstep4 \font\titleis=cmmi7 scaled\magstep4 {
\Large{\textbf{Entropic destruction of a moving heavy quarkonium}
\\}}
\vspace{1.5cm} \noindent{{
Kazem Bitaghsir
Fadafan$^{a}$\footnote{e-mail:bitaghsir@shahroodut.ac.ir }, Seyed
Kamal Tabatabaei$^{b}$\footnote{e-mail:k.tabatabaei67@yahoo.com }
}}\\
\vspace{0.8cm}
{\em
${^a}$Physics Department, Shahrood University of Technology, \\Shahrood, Iran\\}
\vspace*{.25cm}
{\em ${^b}$ School of Physics, Institute for Research in Fundamental
Sciences (IPM), P.O.Box 19395-5531, Tehran, Iran}


\vspace*{.4cm}

\end{center}
\vskip 2em

\begin{abstract}
Recently it has been shown that the peak of the quarkonium entropy at the deconfinement transition is related to the emergent entropic force which destructs the quarkonium. Using the AdS/CFT correspondence, we consider dissociation of a moving heavy quarkonium by entropic force. We find that the entropic force destructs the moving quarkonium easier than the static case which is expected from perturbative weakly coupled plasma. By considering the Maxwell charge, we study the effect of medium on the destruction of heavy quarkonium. It is shown that the quarkonium dissociates easier in the medium.

\end{abstract}

\end{titlepage}

\section{Introduction}
In heavy ion collisions at LHC and RHIC, one of the main experimental signatures of the
formation of a strongly coupled quark$-$gluon plasma (QGP) is melting of heavy quarkoniums \cite{Matsui:1986dk}. They are expected to produce during the initial stages of the collision then give us significant information about the entire evolution of the QGP. For example, it was predicted that the heavy quarks of charm $(c)$ and anti-charm $(\bar{c})$, i.e. charmonium, is suppressed due to the Debye screening effects induced by the high density of color charges in the QGP. It was shown that at RHIC the charmonium suppression is stronger than at LHC while the density is larger at LHC \cite{Adare:2006ns,Abelev:2013ila}. This is known as the puzzle on the suppression of the charmonium at RHIC and LHC. On the other hand, the melting of heavy quarkoniums are related to the deconfinement transition and Lattice QCD implies that there is a peak in the entropy of heavy quarkonium at the deconfinement transition. Recently, it was argued that the puzzle on the suppression of the charmonium is related to the nature of deconfinement \cite{Kharzeev:2014pha}. This argument is based on the Lattice results which indicate a large amount of entropy associated with the heavy quark-antiquark pair placed in the QGP \cite{Kaczmarek:2002mc,Petreczky:2004pz}.

In the proposal of \cite{Kharzeev:2014pha}, the entropy $S$ which grows with the inter-quark distance $L$ leads to the entropic force
 \be
 \mathcal{F}=T\frac{\partial S}{\partial L}.
  \ee
here, $T$ is the temperature of the plasma. This entropic force would be solution of the puzzle and is responsible to dissociate the quarkonium.

The entropic force is an emergent force and does not describe other fundamental interactions. Based on the second low of thermodynamics, it originates from multiple interactions driving the system towards the state with a larger entropy. It was originally introduced in \cite{entropic1} and proposed by Verlinde that the entropic force may be responsible for gravity \cite{Verlinde:2010hp}. We will not discuss on this intriguing idea and restrict
ourselves to its application in melting of quarkonium in the QGP.

Using the AdS/CFT correspondence is a new method for studying different aspects of the QGP \cite{CasalderreySolana:2011us}. This method has yielded many important insights into the dynamics of strongly coupled gauge theories.
In this approach, the suppression of static charmonium has been studied in \cite{Hashimoto:2014fha} and it was shown that the peak of the entropy emerges when the U-shaped string stretched between the heavy quark and antiquark touches the horizon of the black hole. From the correspondence, it means that in the boundary theory, the condensation of long strings will span the entire volume of the finite temperature system \cite{Hashimoto:2014xta}.

In this paper, we extend the holographic studies of \cite{Hashimoto:2014fha} to the case of a moving charmonium. Because in heavy ion collisions, charmonium which observed experimentally is moving with relativistic velocities through the QGP. We address behavior of moving charmonium at finite temperature and density and study how the deconfinement transition can be viewed as entropic self-destruction. We have studied before the effect of finite 't Hooft coupling corrections on the thermal width
and imaginary potential of a moving heavy charmonium from holography in \cite{Fadafan:2015kma}. It was argued that by
considering such corrections the thermal width becomes effectively smaller. It was found that the moving quarkonium dissociate easier than the static case in agreement with the QCD expectations \cite{Ali-Akbari:2014vpa,Finazzo:2014rca}.

First, we study the moving charmonium within two different theories: $N = 4$ supersymmetric Yang-Mills (SYM) theory where at zero temperature possesses conformal invariance, and pure Yang-Mills (YM) theory that possesses confinement-deconfinement phase transition at a critical temperature. We extend this study to the confinement/deconfinement phase transition in the presence of quark medium \cite{Kim:2007em}. In this case, we study the entropic force in the medium composed of light quarks and gluons. At low temperature, the dual gravity to the hadronic phase is the thermal
charged AdS space time. At high temperature, the geometry dual to the QGP is the AdS Reissner-Nordstrom AdS $(AdSRN)$ black hole. The the jet quenching parameter and the energy loss of a moving heavy quark in the background of AdSRN black hole are studied in \cite{Fadafan:2008uv}. Also melting of a heavy quarkonium in this background has been investigated in \cite{Fadafan:2012qy}. It was found  there is a same mechanism for melting of quarkonium in the hadronic phase and in the QGP, i.e. the interaction between heavy quark and antiquarks is screened by the light quarks.

To study the moving quarkonium, it would be possible to consider different alignments for charmonium with respect to the plasma wind. They are parallel ($\theta=0$), transverse ($\theta=\pi/2$) or arbitrary direction to the wind ($\theta$). We study only parallel and transverse cases in this paper.

This paper is organized as follows. In the next section, we will present the general expressions to study the holographic entropic destruction of a moving quarkonium for ($\theta=\pi/2$). For ($\theta=0$), we concern mainly on the final results and show related plots. In section three, we consider confining backgrounds. We study the entropy of the moving quarkonium in the presence of medium and in confining theory in section four. In the last section we will summarize our final results. %

\section{The entropic force of a moving quarkonium from holography}
We begin by using the AdS/CFT correspondence to calculate the additional entropy associated with adding a
color singlet heavy charmonium to the QGP. We assume the general background as follows:
\be
 ds^2=G_{tt}dt^2+G_{xx}dx_i^2+G_{uu}du^2,\label{general-background}
\ee%
here the metric elements are functions of the radial distance $u$
and $x_i=x,y,z$ are the boundary coordinates. In these coordinates,
the boundary is located at infinity. From the AdS/CFT correspondence, the coupling constant which is denoted as 't Hooft coupling
$\lambda$ is related to the curvature radius of the $AdS_5$ and
$S_5$ $(R)$, also the string tension $(\frac{1}{2\pi \alpha'})$ by
$\sqrt{\lambda}=\frac{R^2}{\alpha'}$.
We also assume that the quark$-$antiquark pair is moving
with rapidity $\beta$. One should notice that in our reference frame
the plasma is at rest. Then we boost the frame in the $z$
direction with rapidity $\beta$ so that $dt = dt' \cosh \beta - dz'
\sinh \beta$ and $dz = -dt' \sinh \beta + dz' \cosh \beta.$ After
dropping the primes, the metric
becomes%
\begin{align}
\label{metricboost} ds^2 = & -\left(|G_{tt}| \cosh^2 \beta - G_{xx}
\sinh^2 \beta\right) dt^2 + \left(G_{xx} \cosh^2 \beta - |G_{tt}|
\sinh^2 \beta\right) dz^2  \nonumber \\ & -2\,\sinh \beta \, \cosh
\beta\left(G_{xx}-|G_{tt}|\right)  \, dt \, dz + G_{xx}
(dx^2+dy^2) + G_{uu} du^2,
\end{align}%

We define new metric functions as follows:%
\begin{subequations}
\label{WVtilde}
\begin{align}
g_1(u) \equiv a_1(u) \cosh^2 \beta - b_1 (u)\sinh^2 \beta \\
g_2(u) \equiv a_2(u) \cosh^2 \beta - b_2 (u)\sinh^2 \beta
\end{align}
\end{subequations}
where $a_1(u)\equiv
-|G_{tt}|G_{uu},\,\,a_2(u) \equiv -|G_{tt}|G_{xx},\,\,\,,\,\,b_1(u)=G_{xx}G_{uu}$ and $b_2(u)=G_{xx}^2.$

One may consider the quark antiquark ($Q\bar{Q}$) pair to be aligned perpendicularly to the wind, in
the $x$ direction %
\be t=\tau,\quad x=\sigma,\quad
u=u(\sigma)~,\label{static-gauge} \ee%
The distance between $Q\bar{Q}$ pair depends on the velocity and
angle, $L(\pi/2,\beta)$, we call it $L$. The quarks are located at
$x=-\frac{L}{2}$ and $x=+\frac{L}{2}$. The deepest point of the U-shaped string is denoted by $u_*\equiv
u(x=0)$. One finds from the action of the U-shaped string
\be\label{staticL1}
L=2\,\int_{u_*}^{\infty}du~\left[
\frac{g_2(u)}{g_1(u)}\left(\frac{g_2(u)}{g_2(u_*)}-1\right)\right]
^{-\frac{1}{2}}.
\ee
The free energy of the $Q\bar{Q}$ pair is equal to the on-shell action of the
fundamental string in the dual geometry. The on-shell action of the connected fundamental string is given by
$F^{(1)}_{Q\bar{Q}}$ as follows:
\bea
F^{(1)}_{Q\bar{Q}}=
\frac{1}{\pi \alpha'}\left[\int_{u_*}^{\infty}du~
\left(\frac{1}{g_1(u)}-\frac{g_2(u_*)}{g_2(u)g_1(u)}\right)^{-1/2}
\right], \label{staticV}
\eea
while for larger $L$, the fundamental string breaks in two disconnected strings. There is an important point for static long U-shaped strings, because it would be possible to add new configurations \cite{Bak:2007fk}. Therefore in the dual gauge theory, the quarks are screened. The free energy in this case is $F^{(2)}_{}$. As it was discussed in \cite{Chernicoff:2006hi}, the choice of $F^{(2)}_{}$ is not unique. To calculate the potential of moving quark and antiquark pair, this free energy should be subtracted from (\ref{staticV}) and choosing different free energies will change the dependency of the potential to the velocity of pair \cite{Natsuume:2007vc}. Following \cite{Chernicoff:2006hi} and \cite{Liu:2006nn}, we consider configuration of two disconnected trailing drag strings \cite{Herzog:2006gh,Gubser:2006bz}. Therefore, the free energy is given by
\bea
F^{(2)}_{}=\frac{1}{\pi \alpha'}\int_{u_h}^{\infty}du. \label{F2}
\eea
The above generic formulas give the related information of the moving heavy quarkonium in terms of the metric elements of a background
\eq{general-background}. In the same manner, one finds the related formulas if considers the $Q\bar{Q}$ pair to be aligned parallel to the wind \cite{Liu:2006he}.

To study the entropic destruction of moving quarkonium, we calculate the entropy as $S=-\frac{\partial F}{\partial T}$. Then for $Q\bar{Q}$ pair and screened quarks, one should calculate $S^{(1)}_{Q\bar{Q}}=-\partial F^{(1)}_{Q\bar{Q}}/\partial T$ and $S^{(2)}=-\partial F^{(2)}/\partial T$, respectively. One finds that $S^{(2)}$, the entropy of moving quarkonium for large distance of $L$, is similar to the static case. One concludes that for isotropic boundary field theory and asymptotic AdS geometries, the entropy of moving quarkonium would be the same as the static case. The important consequence of this observation is related to the the peak of the quarkonium entropy at the deconfinement transition. We find that this peak exists also for dynamical quarkonium. This observation confirms that the peak is related to the nature of confinement/deconfinement transition and is generic future of such theories. It would be interesting to check this result from Lattice QCD.

In the following, we consider N = 4 SYM theory to study behavior of quarkonium's entropy.
Using the AdS/CFT correspondence, the dual metric functions to N=4 SYM are %
\be
a_2(u)=\frac{u^4-u_h^4}{R^4},\,\,\,\,\,a_1(u)=1,\,\,\,b_2(u)=\frac{u^4}{R^4}
,\,\,\,b_1(u)=\left(1-\frac{u_h^4}{u^4}\right)^{-1}.
\ee%
where the horizon is located at $u_h$ and the temperature of the hot
plasma is given by $u_h=\pi R^2 T$. By comparing $F^{(1)}_{Q\bar{Q}}$ and $F^{(2)}_{}$, one finds numerically if $L> \frac{c}{T}$ the quarks are completely screened where numerically the constant is found as $c=0.28$. In this case
$ S^{(2)}_{}=\sqrt{\lambda}\,\theta \left(L-\frac{c}{T}\right)$.\footnote{Our notation differs from \cite{Hashimoto:2014fha}. }

\begin{figure}
\centerline{\includegraphics[width=3in]{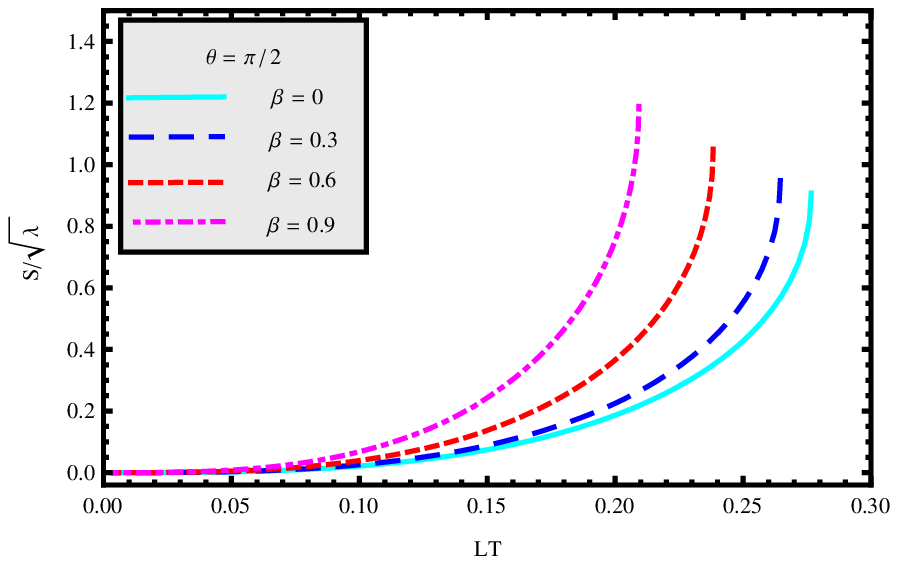},\includegraphics[width=3in]{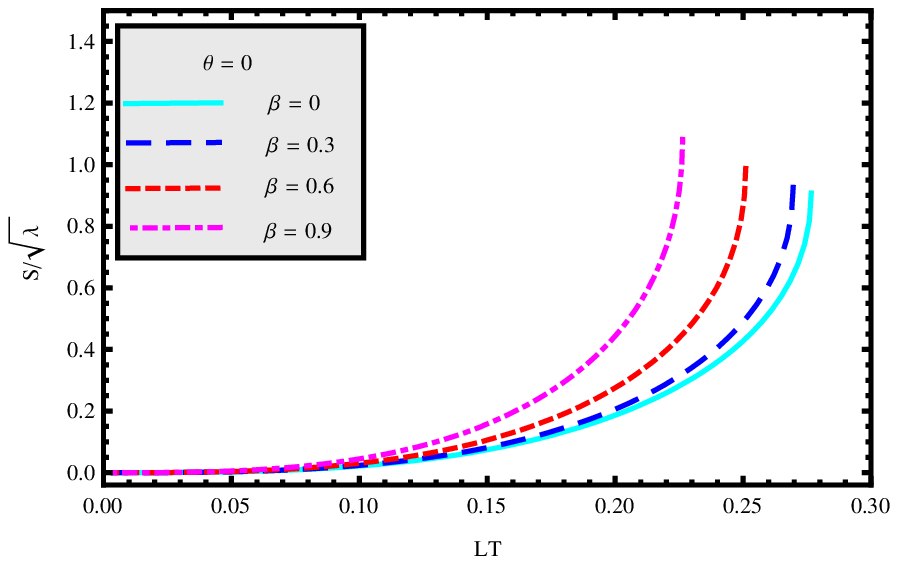}}%
\caption{The entropy of the moving heavy quarkonium in the $N=4$ SYM versus $LT$
 when the quarkonium is moving transverse (left plot) and parallel (right plot) to the wind. From top to bottom the velocity is $\beta=0.9,0.6,0.3,0$.
}\label{N490}
\end{figure}

For $L< \frac{c}{T}$, we show the growth of the entropy $S^{(1)}_{Q\bar{Q}}$ with the inter-quark distance in Fig. 1. In the left plot of this figure, the quarkonium is moving transverse to the wind while in the right plot is moving parallel to the wind.  As it is clear in this figure, increasing the velocity of quark leads to stronger entropy at small distances. One should notice from (1) that the growth of the entropy with the distance is responsible for the entropic force. This force leads to destruction of the quarkonium. Because of increasing entropy at small distances, one concludes that the quarkonium will dissociate easier by increasing the velocity. Briefly, nonzero velocity leads to increase of the entropy at smaller distance of $Q\bar{Q}$ pair. In Fig. 2, we show that the enhancement is stronger when the quarkonium moves orthogonal to the QGP wind. Based on these results one finds that the moving quarkonium dissociates easier than the static ones in agreement with the experimental expectations.

\begin{figure}
\centerline{\includegraphics[width=3in]{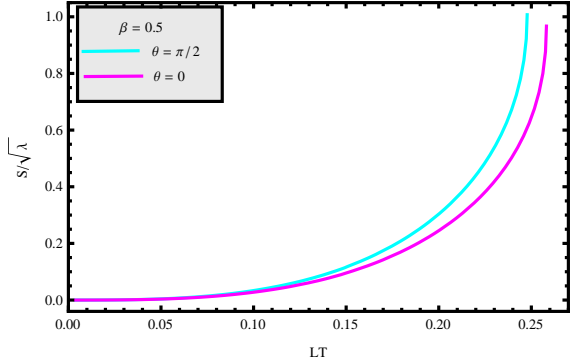}}%
\caption{ In this figure, we compare the entropy of the moving heavy quarkonium transverse (top curve) and parallel (bottom curve) in the $N=4$ SYM versus $LT$.
 }\label{N490}
\end{figure}
However, in $N = 4$ SYM one finds a temperature-independent value for the entropy at large distances.
\section{Entropy of a moving quarkonium in confining theory}
To study entropic force behavior around the deconfinement transition, one should consider a theory which at zero temperature exhibits a confinement phase. For larger separation of quark and antiquark, the linear potential grows up and the string tension should be increased. This study has been done in the confining geometries  from holography in \cite{Giataganas:2014mla}. From holography the free energy at large distance has been found in \cite{Hashimoto:2014fha}.
For example, one may consider the confining $SU(N)$ gauge theory based on $N D4$ branes on a circle \cite{Witten:1998zw}. Two fundamental parameters are a energy scale and temperature. In the vacuum of the theory at zero temperature, the color charge is confined and it exhibits linear confinement of quarks. The gravitational geometry dual to the vacuum is known
analytically as %
\bea
ds^2&=&(\frac{u}{l})^{3/2}\left(-dt^2+d\rho^2+\rho^2d\phi^2+dx_3^2+h(u)dx_4^2\right)+(\frac{l}{u})^{3/2}
\left(\frac{du^2}{h(u)}+u^2d\Omega_4^2\right) \nonumber\\
&& F_{(4)}=\frac{2\pi N}{V_4}\epsilon_4,\,\,\,\,\,e^{\phi}=g_s(\frac{u}{l})^{3/4},\,\,\,\,l^3=\pi g_s N_c l_s^3,\,\,\,\,h(u)=1-(\frac{u_k}{u})^3.\label{LowT}
\eea%
Here all dimensionful quantities measure in units of $l$ which is a length scale related to the D4 brane. $V_4$
and $\epsilon_4$ are volume of the unit $S^4$ and the associated volume form, respectively. Also
$ \frac{u_k}{l}=\frac{4l^2}{9R^2}$ demanding absence of a conical
singularity at the tip of the cigar $u_k$ that is spanned by $u$ and $x_4$. In the deconfined phase, one should consider the black hole geometry as
\bea
ds^2&=&(\frac{u}{l})^{3/2}\left(-H(u)dt^2+d\rho^2+\rho^2d\phi^2+dx_3^2+dx_4^2\right)+(\frac{l}{u})^{3/2}
\left(\frac{du^2}{H(u)}+u^2d\Omega_4^2\right). \label{HighT}\eea%
here $H(u)$ and the temperature are given by%
\be H(u)=1-(\frac{u_h}{u})^3,\,\,\,\,\,\frac{u_h}{l}=\frac{16\pi^2 l^2}{9}T^2. \ee%
The critical temperature is $T_c=\frac{1}{2\pi R}$. The
theory for $T>T_c$ is deconfined and is
given by \eq{HighT}, whereas, for $T<T_c$ is confined and the gravity dual is described by
\eq{LowT}. We have studied the entropy of the moving charmonium transverse and parallel to the wind in this background. We find the same behavior for entropy as shown in figures (1) and (2). The important behavior of the entropy in confining geometries is shown in Fig. 3. As it was discussed in \cite{Hashimoto:2014fha}, this behavior captures the characteristic behavior of the entropy which was found in lattice QCD \cite{Kaczmarek:2005gi}. We have discussed before that for moving or static quarkonium this behavior does not change.

Recently, the real-time dynamics of the dissociation of the quarkonium is studied by considering
the holographic classical string falling in the black hole horizon \cite{Iatrakis:2015sua}. Melting in this case means that for larger distances $L$, the gluon cloud around the quark and antiquark pair will eventually thermalize and become part of the plasma medium.  Another possible mechanism for melting of the quarkonium is imaginary potential between quarks. By studying the perturbations of the U-shaped string near the horizon of the black hole, one can produce the imaginary part of the quarkonium potential which leads to melting of it. We argued in \cite{Fadafan:2013coa} that if $u_k > u_h$ then the connected U-shaped string cannot go past $u_k$ and one cannot consider such fluctuations beyond $u_k$. Then, for $T<T_c$ the imaginary potential vanishes while for $T>T_c$ it is not zero. Hence, we argue that one important mechanism for melting of heavy quarkonium in this setup would be the imaginary potential. It would be interesting to compare the competition between entropic force and imaginary potential for melting of mesons in this case. We leave this problem for further study.
\begin{figure}
\centerline{\includegraphics[width=3in]{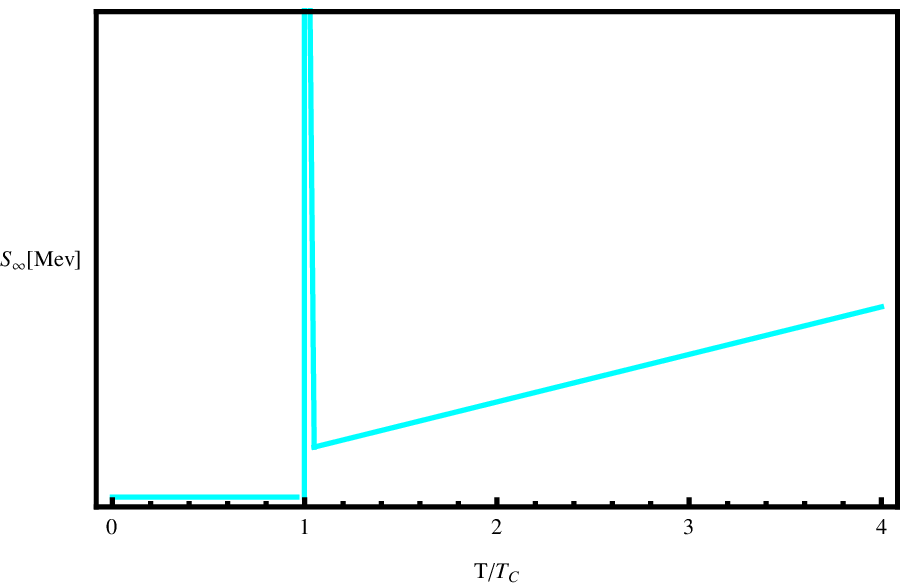}\includegraphics[width=3in]{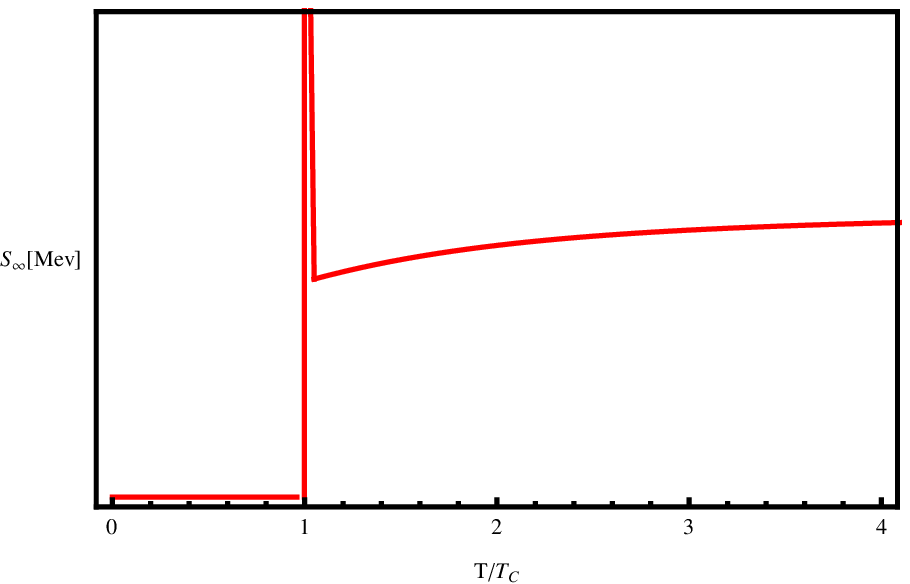}}%
\caption{ The entropy of the quark-antiquark pair at large distance in the confining
theories as a function of $T/T_c$. Left: in the confining YM theory. Right: at finite density with $\mu=30MeV$ and $n=2/3$. }\label{mentr}
\end{figure}
\section{Medium effect on the entropic force }
In this section, we consider the confinement/deconfinement phase transition in the presence of quark medium and study how the medium composed of light quarks and gluons affect the entropic force. Two different phases are hadronic and QGP. At low temperature, the dual gravity to the hadronic phase is the thermal charged AdS spacetime. At high temperature, the geometry dual to the QGP is the $AdSRN$ black hole. Recently from extremal AdSRN black holes, the drag force in a strongly-coupled boundary field theory at finite charge density and zero temperature has been studied in \cite{Ahmadvand:2015gfi}.

The Euclidean action which describes the five-dimensional asymptotic AdS spacetime with the gauge field is given by
\be S = \int d^5 x \sqrt{G} \left( \frac{1}{2 \kappa^2} \left( -
{\cal R} + 2 \Lambda\right)  + \frac{1}{4g^2} F_{MN} F^{MN} \right)
. \label{S1}\ee%
Here $\kappa^2$ and $g^2$ are proportional to the five-dimensional Newton
constant and the five-dimensional gauge coupling constant, respectively.
The cosmological constant is also given by $\Lambda = \frac{-6}{R^2}$. From the AdS/CFT correspondence, the density in the dual field theory is mapped to a bulk gauge field.

Here, we consider the high temperature phase which is deconfined phase and denoted as QGP. The gravity dual is given by
\be ds^2= R^2 u^2 \left( - p(u) dt^2 + d \vec{x}^{ 2} + \frac{1}{u^4
p(u)} du^2 \right)  ,
\label{ds2}\ee%
here $p(u)$ is given by
\be p(u)=1-m u^{-4}+ q^2 u^{-6}.\label{fQGP}\ee%
As before, the boundary is located at
infinity and the geometry is asymptotically $AdS$ with radius $R$.
The black hole temperature $(T)$ and mass $(m)$ are given by %
 \be m=u_h^4+q^2u_h^{-2},\,\,\,\,\,\,T=\frac{u_h}{\pi} \left( 1- \frac{q^2}{2 u_h^6}\right).\ee
where $q$ is the black hole charge. The time-component of the bulk gauge field is
$A_t(u)=i(2\pi^2\mu-Q\,u^{-2})$ where $\mu$ and $Q$ are associated to
the chemical potential and quark number density in the dual field
theory. One finds from the Drichlet boundary condition at the horizon,
$A_t(u_h)=0$, and $Q=2 \pi^2 \mu \,u_h$. Also $ Q=\sqrt{\frac{3}{2n}}\,q$
where $n^{-1}=\frac{g^2R^2}{\kappa^2}$ is the color number. It is defined as ration of number of flavors to number of colors $n=\frac{N_F}{N_C}$.
\begin{figure}
\centerline{\includegraphics[width=3in]{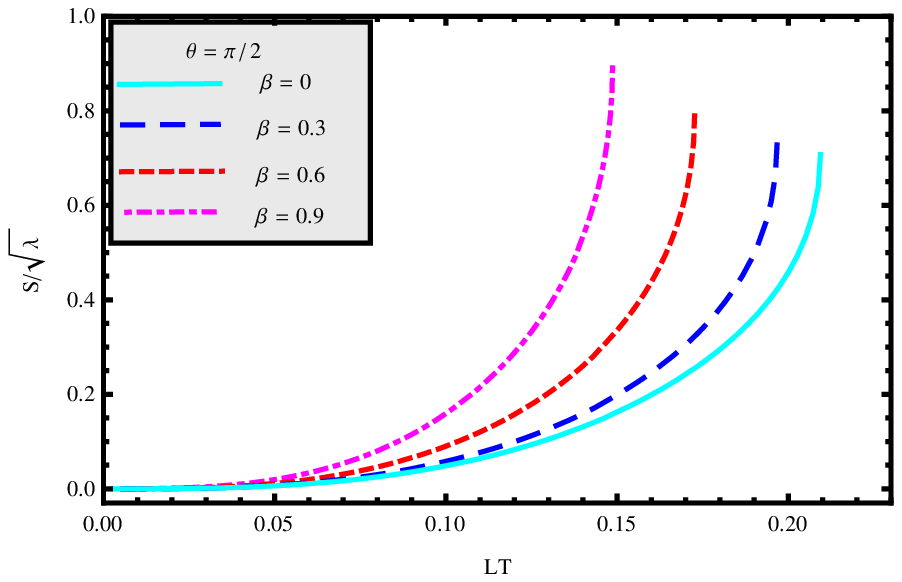},\includegraphics[width=3in]{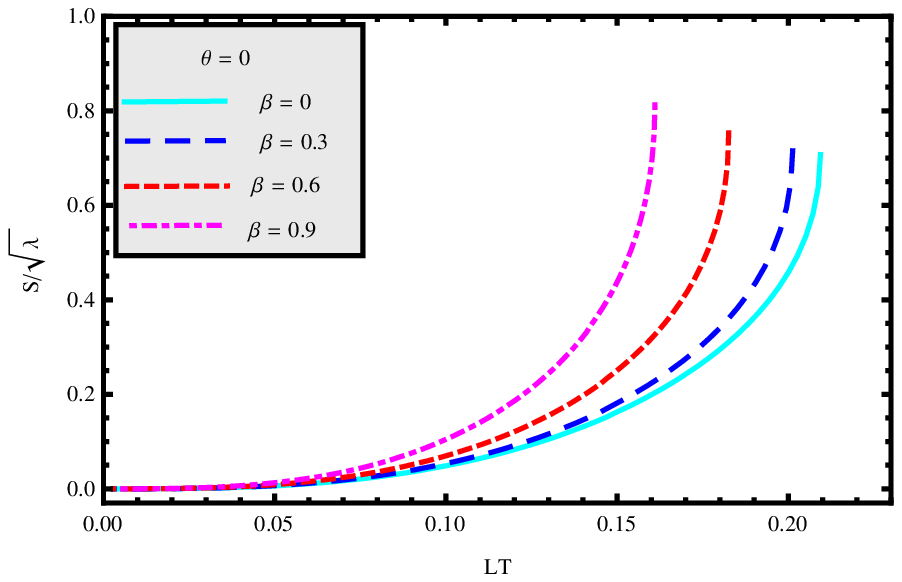}}%
\caption{The effect of medium on the entropy of the moving heavy quarkonium in the medium, with $n=1$ and $\mu=55MeV$, versus $LT$
 when the quarkonium is moving transverse (left plot) and parallel (right plot) to the wind. From top to bottom the velocity is $\beta=0.9,0.6,0.3,0$.
}\label{N490}
\end{figure}

The behavior of the entropy versus the quark and antiquark distance in the confining YM theory and finite density are shown in left and right plots of Fig. 3, respectively. They are qualitatively the same as each other. However, at large distance the dependency of the entropy to the temperature of the plasma differs. The medium of light quarks and gluons is responsible for this behavior. It would be interesting to check this result from Lattice QCD, too. Although the model is not the real QCD and many assumptions have been considered to simplify the model, it captures the characteristic behavior of the entropy which was reported in lattice QCD.

We have also studied the effect of the medium on the entropy in Fig. 4. In the left plot of this figure, the quarkonium is moving transverse to the wind. One finds the same physics as before, \emph{i.e} by increasing the velocity of the quarkonium the maximum of the entropy happens at smaller distance. The same behavior can be seen from the right plot of this figure when the quarkonium is moving parallel to the wind. One finds from Fig. 5 that the entropy of moving quarkonium transverse to the wind is larger than the parallel case. Again Lattice QCD results should be compared to understand the physics.

To understand the effect of the finite density on the entropy, we compare Figs 1 and 5. It is clear that in the presence of medium, growing of entropy happens at smaller distance. Then one concludes that the moving quarkonium dissociates easier at finite density.

\begin{figure}
\centerline{\includegraphics[width=3in]{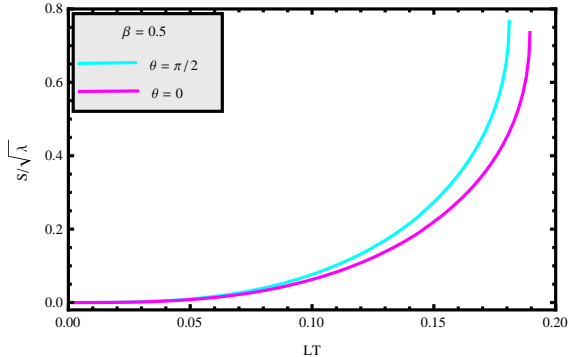}}%
\caption{ The entropy of the moving heavy quarkonium at finite density versus $LT$
 when the quarkonium is moving transverse (top curve) and parallel (bottom curve) to the wind.
}\label{N490}
\end{figure}
\section{Conclusion}
In this paper, we have studied destruction of a moving heavy quarkonium due to the entropic force from the AdS/CFT correspondence. In heavy ion collisions at RHIC and LHC, most of quarkonium observed experimentally are moving through the QGP with relativistic velocities. Because of the well known result that the QGP near the transition temperature is a strongly coupled system, using the AdS/CFT correspondence would be reliable. Using this method, the entropy of heavy quarkonium was studied in \cite{Hashimoto:2014fha}. It was shown that for static quarkonium the strong peak in the entropy $S$ near the transition point is related to the nature of deconfinement. From holographic point of view, $S$ originates from a long string in the confining theory where at the deconfinement transition point is absorbed by a black hole. As it was discussed this peak is absent in the conformal gauge theory like N=4 SYM but emerges in a confining theory. One finds that the entropic force resulting from the distribution of heavy quarks is  responsible for melting of meson.

We discussed that another possible mechanism for meson melting would be the imaginary part of the potential between quark and antiquark. The entropic force is responsible for the diffusion of quarks in the plasma and it would be very interesting to study effect of different velocities of quark and antiquark for meson melting.

It was shown that this peak exists also for moving quarkonium. It would be interesting to study this phenomenon from perturbative QCD or Lattice techniques. This observation confirmed that the peak of the moving quark
pair entropy at deconfinement transition is a generic feature of the
holographic confining models.

We studied this phenomenon in N = 4 SYM theory where the theory does not possess a deconfinement transition and pure confining YM theory and AdSRN background black hole where the theory at $T = 0$ possesses confinement. To study the charge effect, we considered Maxwell charge which can be interpreted as quark medium. As the case of static quarkonium, we find that the entropy grows with increasing the inter-quark distance. We conclude that the quarkonium dissociates easier in the medium.



\begin{thebibliography}{99}
\bibitem{Matsui:1986dk}
  T.~Matsui, H.~Satz,
  J/psi Suppression by Quark-Gluon Plasma Formation.
  Phys.\ Lett.\ B {\bf 178} (1986) 416.
\bibitem{Adare:2006ns}
  A.~Adare {\it et al.} [PHENIX Collaboration],
  $J/\psi$ Production vs Centrality, Transverse Momentum, and Rapidity in Au+Au Collisions at $\sqrt{s_{NN}} = 200$ GeV
  Phys.\ Rev.\ Lett.\  {\bf 98} (2007) 232301
  [nucl-ex/0611020].
\bibitem{Abelev:2013ila}
  B.~B.~Abelev {\it et al.} [ALICE Collaboration],
  Centrality, rapidity and transverse momentum dependence of $J/\psi$ suppression in Pb-Pb collisions at $\sqrt{s_{\rm NN}}$=2.76 TeV
  Phys.\ Lett.\ B {\bf 734} (2014) 314
  [arXiv:1311.0214 [nucl-ex]].

\bibitem{Kharzeev:2014pha}
  D.~E.~Kharzeev,
  Deconfinement as an entropic self-destruction: a solution for the quarkonium suppression puzzle?
  Phys.\ Rev.\ D {\bf 90} (2014) 7,  074007
  [arXiv:1409.2496 [hep-ph]].

\bibitem{Kaczmarek:2002mc}
  O.~Kaczmarek, F.~Karsch, P.~Petreczky and F.~Zantow,
  Heavy quark anti-quark free energy and the renormalized Polyakov loop
  Phys.\ Lett.\ B {\bf 543} (2002) 41
  [hep-lat/0207002].
\bibitem{Kaczmarek:2005gi}
  O.~Kaczmarek and F.~Zantow,
  ``Static quark anti-quark interactions at zero and finite temperature QCD. II. Quark anti-quark internal energy and entropy,''
  hep-lat/0506019.
\bibitem{Petreczky:2004pz}
  P.~Petreczky and K.~Petrov,
  Free energy of a static quark anti-quark pair and the renormalized Polyakov loop in three flavor QCD
  Phys.\ Rev.\ D {\bf 70} (2004) 054503
  [hep-lat/0405009].
\bibitem{entropic1}
K.~H.~ Meyer, G.~Susich, and E.~ Valk,
Elastic properties of rubber-like substances
 Kolloid~Z. 59, 208
(1932).
\bibitem{Verlinde:2010hp}
  E.~P.~Verlinde,
  On the Origin of Gravity and the Laws of Newton
  JHEP {\bf 1104} (2011) 029
  [arXiv:1001.0785 [hep-th]].
\bibitem{CasalderreySolana:2011us}
  J.~Casalderrey-Solana, H.~Liu, D.~Mateos, K.~Rajagopal, U.~A.~Wiedemann,
  Gauge/string duality, hot QCD and heavy ion collisions.
  arXiv:1101.0618 [hep-th].


\bibitem{Hashimoto:2014fha}
  K.~Hashimoto and D.~E.~Kharzeev,
  Entropic destruction of heavy quarkonium in non-Abelian plasma from holography
  Phys.\ Rev.\ D {\bf 90} (2014) 12,  125012
  [arXiv:1411.0618 [hep-th]].
\bibitem{Fadafan:2015kma}
  K.~B.~Fadafan and S.~K.~Tabatabaei,
  ``The Imaginary Potential and Thermal Width of Moving Quarkonium from Holography,''
  arXiv:1501.00439 [hep-th].
\bibitem{Ali-Akbari:2014vpa}
  M.~Ali-Akbari, D.~Giataganas and Z.~Rezaei,
  ``Imaginary potential of heavy quarkonia moving in strongly coupled plasma,''
  Phys.\ Rev.\ D {\bf 90} (2014) 8,  086001
  doi:10.1103/PhysRevD.90.086001
  [arXiv:1406.1994 [hep-th]].
\bibitem{Finazzo:2014rca}
  S.~I.~Finazzo and J.~Noronha,
  ``Thermal suppression of moving heavy quark pairs in a strongly coupled plasma,''
  JHEP {\bf 1501} (2015) 051
  doi:10.1007/JHEP01(2015)051
  [arXiv:1406.2683 [hep-th]].
\bibitem{Hashimoto:2014xta}
  K.~Hashimoto, S.~Kinoshita, K.~Murata and T.~Oka,
  Turbulent meson condensation in quark deconfinement
  Phys.\ Lett.\ B {\bf 746} (2015) 311
  [arXiv:1408.6293 [hep-th]].
\bibitem{Bak:2007fk}
  D.~Bak, A.~Karch, L.~G.~Yaffe,
  Debye screening in strongly coupled N=4 supersymmetric Yang$-$Mills
  plasma.
  JHEP {\bf 0708} (2007) 049
  [arXiv:0705.0994 [hep-th]].
\bibitem{Chernicoff:2006hi}
  M.~Chernicoff, J.~A.~Garcia and A.~Guijosa,
  ``The Energy of a Moving Quark-Antiquark Pair in an N=4 SYM Plasma,''
  JHEP {\bf 0609}, 068 (2006)
  doi:10.1088/1126-6708/2006/09/068
  [hep-th/0607089].
\bibitem{Natsuume:2007vc}
  M.~Natsuume and T.~Okamura,
  ``Screening length and the direction of plasma winds,''
  JHEP {\bf 0709}, 039 (2007)
  doi:10.1088/1126-6708/2007/09/039
  [arXiv:0706.0086 [hep-th]].
\bibitem{Liu:2006nn}
  H.~Liu, K.~Rajagopal and U.~A.~Wiedemann,
  ``An AdS/CFT Calculation of Screening in a Hot Wind,''
  Phys.\ Rev.\ Lett.\  {\bf 98}, 182301 (2007)
  doi:10.1103/PhysRevLett.98.182301
  [hep-ph/0607062].

 \bibitem{Herzog:2006gh}
  C.~P.~Herzog, A.~Karch, P.~Kovtun, C.~Kozcaz and L.~G.~Yaffe,
  ``Energy loss of a heavy quark moving through N=4 supersymmetric Yang-Mills plasma,''
  JHEP {\bf 0607} (2006) 013
  [hep-th/0605158].
\bibitem{Gubser:2006bz}
  S.~S.~Gubser,
  ``Drag force in AdS/CFT,''
  Phys.\ Rev.\ D {\bf 74} (2006) 126005
  [hep-th/0605182].


\bibitem{Giataganas:2014mla}
  D.~Giataganas and K.~Goldstein,
  ``Tension of Confining Strings at Low Temperature,''
  JHEP {\bf 1502}, 123 (2015)
  doi:10.1007/JHEP02(2015)123
  [arXiv:1411.4995 [hep-th]].
\bibitem{Liu:2006he}
  H.~Liu, K.~Rajagopal, U.~A.~Wiedemann,
  Wilson loops in heavy ion collisions and their calculation in
  AdS/CFT.
   JHEP {\bf 0703} (2007) 066  [hep-ph/0612168].
\bibitem{Witten:1998zw}
  E.~Witten,
  Anti-de Sitter space, thermal phase transition, and confinement in  gauge
  theories.
  Adv.\ Theor.\ Math.\ Phys.\  {\bf 2} (1998) 505
  [arXiv:hep-th/9803131].
\bibitem{Kim:2007em}
  Y.~Kim, B.~-H.~Lee, S.~Nam, C.~Park and S.~-J.~Sin,
  ``Deconfinement phase transition in holographic QCD with matter,''
  Phys.\ Rev.\ D {\bf 76} (2007) 086003
  [arXiv:0706.2525 [hep-ph]].
\bibitem{Iatrakis:2015sua}
  I.~Iatrakis and D.~E.~Kharzeev,
  ``Holographic entropy and real-time dynamics of quarkonium dissociation in non-Abelian plasma,''
  arXiv:1509.08286 [hep-ph].
\bibitem{Fadafan:2013coa}
  K.~B.~Fadafan and S.~K.~Tabatabaei,
  ``Thermal Width of Quarkonium from Holography,''
  Eur.\ Phys.\ J.\ C {\bf 74} (2014) 2842
  [arXiv:1308.3971 [hep-th]].
\cite{Fadafan:2008uv}
\bibitem{Fadafan:2008uv}
  K.~B.~Fadafan,
  ``Charge effect and finite 't Hooft coupling correction on drag force and Jet Quenching Parameter,''
  Eur.\ Phys.\ J.\ C {\bf 68} (2010) 505
  [arXiv:0809.1336 [hep-th]].
\bibitem{Fadafan:2012qy}
  K.~B.~Fadafan and E.~Azimfard,
  ``On meson melting in the quark medium,''
  Nucl.\ Phys.\ B {\bf 863}, 347 (2012)
  doi:10.1016/j.nuclphysb.2012.05.022
  [arXiv:1203.3942 [hep-th]].
\bibitem{Ahmadvand:2015gfi}
  M.~Ahmadvand and K.~B.~Fadafan,
  ``Energy loss at zero temperature from extremal black holes,''
  arXiv:1512.05290 [hep-th].



\end{thebibliography}
\end{document}